\documentclass[preprint,preprintnumbers, prd, floatfix, superscriptaddress,nofootinbib] {revtex4-1}
\usepackage{epsfig}
\usepackage{subfigure}
\usepackage{dcolumn}
\usepackage{bm}
\usepackage[usenames ,dvipsnames]{xcolor}
\usepackage{slashed}
\usepackage{graphicx,color}
\usepackage{ulem}
\begin{document}
\title{A diagrammatic analysis of
two-body charmed baryon decays
with flavor symmetry
}

\author{H.J. Zhao}
\email{hjzhao@163.com}
\affiliation{School of Physics and Information Engineering, Shanxi Normal University, Linfen 041004, China}

\author{Yan-Li Wang}
\email{1556233556@qq.com}
\affiliation{School of Physics and Information Engineering, Shanxi Normal University, Linfen 041004, China}

\author{Y.K. Hsiao}
\email{yukuohsiao@gmail.com}
\affiliation{School of Physics and Information Engineering, Shanxi Normal University, Linfen 041004, China}

\author{Yao Yu}
\email{yuyao@cqupt.edu.cn}
\affiliation{Chongqing University of Posts \& Telecommunications, Chongqing, 400065, China}

\date{\today}

\begin{abstract}
We study the two-body anti-triplet charmed baryon decays
based on the diagrammatic approach with $SU(3)$ flavor symmetry.
We extract
the two $W$-exchange effects as $E_{\bf B}$ and $E^\prime$
that contribute to the $\Lambda_c^+ \to \Xi^0 K^+$ decay,
together with the relative phases,
where $E_{\bf B}$ gives the main contribution.
Besides,
we find that
${\cal B}(\Lambda_c^+\to p\pi^0)=(0.8 ^{+0.9}_{-0.8})\times 10^{-4}$,
which is within the experimental upper bound.
Particularly, we obtain
${\cal B}(\Xi_c^+\to\Xi^{0} \pi^{+})=(9.3 \pm 3.6)\times 10^{-3}$,
${\cal B}(\Xi_c^0\to \Xi^-\pi^+,\Lambda^0\bar K^0)
=(19.3 \pm 2.8,8.3 \pm 5.0)\times 10^{-2}$ and
${\cal B}(\Xi_c^0\to \Xi^- K^+)=(5.6 \pm 0.8)\times 10^{-4}$,
which all agree with the data.
For the singly Cabibbo suppressed $\Lambda_c^+$ decays,
we predict that
${\cal B}(\Lambda_c^+ \to n \pi^{+},p \eta^\prime,\Sigma^{+} K^{0})
=(7.7\pm 2.0,7.1 \pm 1.4,19.1 \pm 4.8)\times 10^{-4}$, which are
accessible to the experiments at BESIII, BELLEII and LHCb.
\end{abstract}

\maketitle

\section{introduction}
The experimental studies of two-body ${\bf B}_c\to {\bf B}M$ decays
have provided important information
for the theoretical understanding of the hadronization in the weak interaction,
where ${\bf B}_c=(\Xi_c^{0},\Xi_c^{+},\Lambda_c^+)$
are the lowest-lying anti-triplet charmed baryon states,
and ${\bf B}(M)$ the baryon (meson) state.
For example,
the BESIII collaboration has recently measured
the purely non-factorizable decays,
of which the branching fractions are given by~\cite{Ablikim:2018bir}
\begin{eqnarray}\label{data_LctoXiK}
{\cal B}(\Lambda_c^+\to \Xi^0 K^+)&=&(5.90\pm 0.86\pm 0.39)\times 10^{-3}\,,\nonumber\\
{\cal B}(\Lambda_c^+\to \Xi^{*0} K^+)&=&(5.02\pm 0.99\pm 0.31)\times 10^{-3}\,,
\end{eqnarray}
with $\Xi^{*0}\equiv \Xi(1530)^0$.
This implies that
the non-factorizable effects can be as significant as the factorizable ones
in ${\bf B}_c\to {\bf B}M$~\cite{Ablikim:2017ors,Ablikim:2015flg},
although being often neglected
in the b-hadron decays~\cite{ali,Geng:2006jt,Hsiao:2014mua,Hsiao:2017tif}.

Some theoretical approaches have tried to
deal with the non-factorizable effects~\cite{XiK_1,XiK_2,XiK_3,XiK_4,XiK_5}.
Nonetheless,
${\cal B}(\Lambda_c^+\to \Xi^0 K^+)$ is calculated to
be 2-6 times smaller than the observation.
Without involving the detailed dynamics,
the approach based on the $SU(3)$ flavor ($SU(3)_f$) symmetry is able to receive
all contributions~\cite{He:2000ys,
Fu:2003fy,Hsiao:2015iiu,He:2015fwa,He:2015fsa,Savage:1989qr,
Savage:1991wu,h_term,Lu:2016ogy,Geng:2017esc,Geng:2018plk,Wang:2017gxe,
Wang:2017azm,Geng:2017mxn,Geng:2018bow,
Geng:2018upx,Hsiao:2019yur,Geng:2018rse,Geng:2019xbo},
such that  ${\cal B}({\bf B}_c\to{\bf B}M)$ can be explained~\cite{Geng:2017esc,
Geng:2018plk,Geng:2017mxn,Geng:2018bow};
particularly, ${\cal B}(\Lambda_c^+\to \Xi^0 K^+)$.
However, the $SU(3)_f$ symmetry mixes 
the factorizable and non-factorizable effects,
instead of quantifying their individual contributions.

\begin{figure}
\centering
\includegraphics[width=1.0
\textwidth]{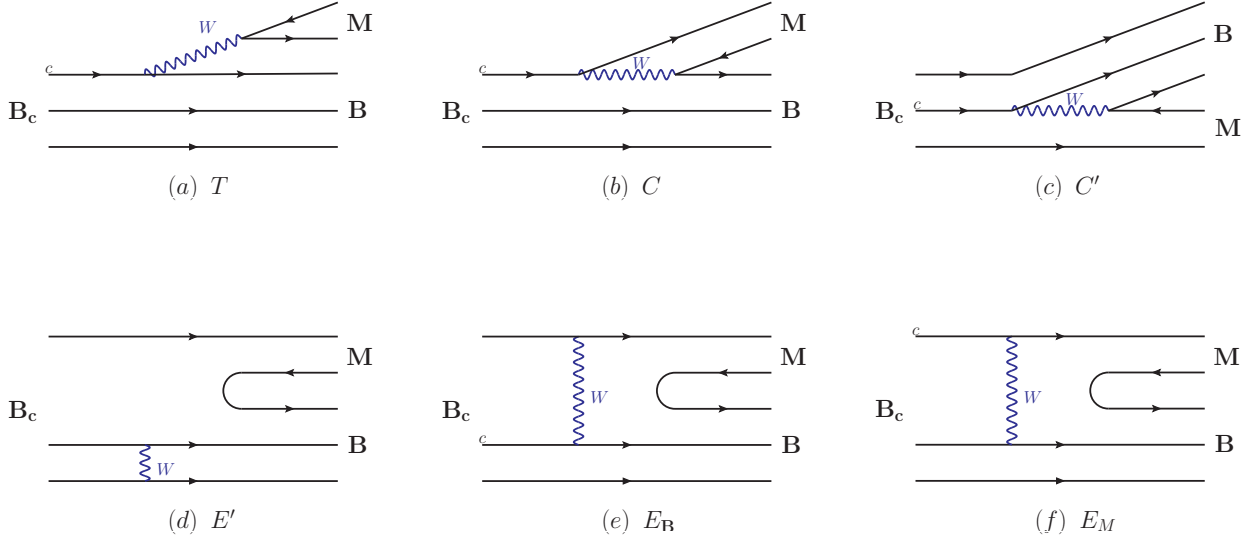} \\
\caption{Topological diagrams for the ${\bf B}_c\to {\bf B}M$ decays.}\label{fig1}
\end{figure}

As depicted in Fig.~\ref{fig1},
one can identify the (non-)factoriable effects
by the topological diagrams, and parameterize them
as the topological amplitudes~\cite{Kohara:1991ug,Chau:1995gk,Cheng:2018hwl}.
Accordingly,
$\Lambda_c^+\to \Xi^{(*)0} K^+$ is seen
to decay through the two $W$-exchange ones in Figs.~\ref{fig1}(d,e).
Since the topological diagrams have been commonly used
in the calculations and measurements~\cite{Ablikim:2018bir,
XiK_1,XiK_2,XiK_3,XiK_4,XiK_5,Cheng:2018hwl,Zou:2019kzq,Leibovich:2003tw},
their information can be important.
Therefore, we propose to perform the numerical analysis with
the topological amplitudes,
such that we can determine the sizes and relative phases
for the different effects.
Note that the same numerical analysis known as the diagrammatic approach
has been well performed to extract the topological amplitudes
in the $D$ decays~\cite{Grossman:2012ry,
Pirtskhalava:2011va,Cheng:2012xb,Li:2012cfa,Li:2013xsa,Cheng:2019ggx}.
In this paper, we will demonstrate that
the topological amplitudes can explain the data.
Particularly, in our fit
we will be able to accommodate
the experimental measurements of 
${\cal B}(\Lambda_c^+\to p\pi^0)<2.7\times10^{-4}$~\cite{Ablikim:2017ors}
and ${\cal B}(\Xi^0_c \to \Xi^- K^+)/{\cal B}(\Xi^0_c \to \Xi^- \pi^+)
=(0.56\pm0.12)s^2_c$~\cite{pdg}, where $s_c=V_{us}$.
We will also predict the branching fractions
for the ${\bf B}_c\to{\bf B}M$ decays that have not been observed yet.
%
%

\section{Diagrammatic approach}
For the two-body charmed baryon decays,
the relevant effective Hamiltonian is given by~\cite{Buras:1998raa}
\begin{eqnarray}\label{Heff}
{\cal H}_{eff}&=&\sum_{i=1,2}\frac{G_F}{\sqrt 2}c_i
\left(V_{cs}V_{ud}O_i+V_{cq}V_{uq} O_i^q+V_{cd}V_{us}O'_i\right),
\end{eqnarray}
with $q=(d,s)$, where $G_F$ is the Fermi constant,
$c_{1,2}$ are the Wilson coefficients, and
$V_{ij}$ the CKM matrix elements.
The four-quark operators $O_{1,2}^{(q)}$ and $O_{1,2}^\prime$ are written as
\begin{eqnarray}\label{O12}
&&
O_1=(\bar u d)(\bar s c)\,,\;O_2=(\bar s d)(\bar u c)\,,\nonumber\\
&&
O_1^q=(\bar u q)(\bar q c)\,,\; O_2^q=(\bar q q)(\bar u c)\,,\nonumber\\
&&
O'_1=(\bar u s)(\bar d c)\,,\;O'_2=(\bar d s)(\bar s c)\,,
\end{eqnarray}
with $q=(d,s)$ and $(\bar q_1 q_2)=\bar q_1\gamma_\mu(1-\gamma_5)q_2$.
The decays with $|V_{cs}V_{ud}|\simeq 1$,
$|V_{cq}V_{uq}|\simeq s_c$ and $|V_{cd}V_{us}|\simeq s_c^2$ are classified as
the Cabibbo-favored (CF), singly Cabibbo-suppressed (SCS)
and doubly Cabibbo-suppressed (DCS) processes, respectively.

By using ${\cal H}_{eff}$ in Eq.~(\ref{Heff}),
we draw different topological diagrams
in ${\bf B}_c\to {\bf B}M$~\cite{Kohara:1991ug,Chau:1995gk,Cheng:2018hwl},
where the quark lines should be in accordance with the operators in Eq.~(\ref{O12}).
As seen in Fig.~\ref{fig1}, we obtain six topological diagrams.
The external and internal $W$-emission diagrams
in Figs.~\ref{fig1}a and b can be parameterized as
the topological amplitudes $T$ and $C$, respectively.
Since one can factorize $T$ and $C$ as
${\cal A}\propto \langle M|(\bar q_1 q_2)|0\rangle
\langle {\bf B}|(\bar q_3 c)|{\bf B}_c\rangle$~\cite{ali},
which consists of two calculable matrix elements,
the $T$ and $C$ are regarded as
the factorizable amplitudes~\cite{Hsiao:2017tif,Geng:2017esc,Cheng:2018hwl}.
The other internal $W$-emission diagram in Fig.~\ref{fig1}c
has no factorizable form, parameterized as $C'$.
In Figs.~1(d,e,f),
the $W$-exchange amplitudes of $(E', E_{\bf B}, E_M)$
need an additional gluon to relate $M$ and ${\bf B}$.
Besides, $E_{\bf B}$($E_M$)
has the $W$-boson to connect ${\bf B}$ and $M$,
with the $c$-quark transition to be a valence quark in ${\bf B}$($M$),
whereas $M$ in the $E'$ amplitude is unable to connect to the $W$-boson.
In addition to $C'$,
$(E', E_{\bf B}, E_M)$ are the non-factorizable amplitudes
according to the factorization approach~\cite{Cheng:2018hwl,Zou:2019kzq}.
As a result,
we clearly identify each (non-)factorizable effect
that contribute to the two-body ${\bf B}_c\to{\bf B}M$ decays.

To present the amplitudes of ${\bf B}_c\to{\bf B}M$ with
$(T,C)$ and $(C',E',E_{\bf B},E_M)$, we need the suitable insertions of the final states
to match the quark lines,
such as $\pi^0=\sqrt{1/2}(u\bar u-d\bar d)$, which adds
a pre-factor of $\pm\sqrt{1/2}$ to the topological parameters.
Likewise,
the $(\eta,\eta')$ meson states that mix with
$\eta_q=\sqrt{1/2}(u\bar u+d\bar d)$ and $\eta_s=s\bar s$
lead to the other pre-factors. Specifically,
the mixing matrix is presented as~\cite{FKS}
\begin{eqnarray}\label{eta_mixing}
\left(\begin{array}{c} \eta \\ \eta^\prime \end{array}\right)
=
\left(\begin{array}{cc} \cos\phi & -\sin\phi \\ \sin\phi & \cos\phi \end{array}\right)
\left(\begin{array}{c} \eta_q \\ \eta_s \end{array}\right),
\end{eqnarray}
with the mixing angle $\phi=(39.3\pm1.0)^\circ$.
We hence obtain the amplitudes of ${\bf B}_c\to{\bf B}M$,
given in Table~\ref{tab1}.
The topological amplitudes are in fact complex,
presented as 11 parameters:
\begin{eqnarray}\label{11p}
&&T, Ce^{i\delta_C},C'e^{i\delta_{C'}},
E_{\bf B}e^{i\delta_{E_{\bf B}}}, E_M e^{i\delta_{E_M}}, E' e^{i\delta_{E'}}\,,
\end{eqnarray}
with $T$ set to be relatively real.
To obtain the decay widths,
we  depend on the integration of the phase space
for the two-body decays, given by~\cite{pdg}
\begin{eqnarray}
&&\Gamma({\bf B}_c\to {\bf B} M)=
\frac{|\vec{p}_{{\bf B}}|}{8\pi m_{{\bf B}_c}^2}|{\cal A}({\bf B}_c\to {\bf B} M)|^2\,,\nonumber\\
&&|\vec{p}_{{\bf B}}|=\frac{\sqrt{[m_{{\bf B}_c}^2-(m_{{\bf B}}+m_M)^2]
[m_{{\bf B}_c}^2-(m_{{\bf B}}-m_M)^2]}}{2 m_{{\bf B}_c}}\,.
\end{eqnarray}
with ${\cal A}({\bf B}_c\to {\bf B}M)$ from Table~\ref{tab1}.

\begin{table}
\caption{Amplitudes of ${\bf B}_c\to {\bf B} M$,
where $\lambda_q\equiv V_{cq}V_{uq}$ with $q=(d,s)$ and
$(s\phi,c\phi)\equiv (\sin\phi,\cos\phi)$ for the $\eta$-$\eta'$ mixing.}\label{tab1}
\tiny
\begin{tabular}{|l|l|}
\hline
Decay modes& Amplitudes/$(\frac{G_F}{\sqrt 2})$ \\
\hline
$\Xi_c^0 \to \Sigma^+ K^-$
&$V_{cs}V_{ud}(E_M+E^\prime)$ \\
$\Xi_c^0 \to \Sigma( \Lambda)^0\bar{K}^0$
&$V_{cs}V_{ud}(C+C^\prime +E_M+E^\prime)$\\
$\Xi_c^0 \to \Xi^0 \pi^0$
&$V_{cs}V_{ud}\frac{1}{\sqrt 2}(E_{\bf B}-C^\prime)$ \\
$\Xi_c^0 \to \Xi^0 \eta $
&$V_{cs}V_{ud}[\frac{1}{\sqrt 2}(C^\prime+E_{\bf B})c\phi-(E_M+E^\prime)s\phi]$\\
$\Xi_c^0 \to \Xi^0 \eta^\prime $
&$V_{cs}V_{ud}[\frac{1}{\sqrt 2}(C^\prime+ E_{\bf B})s\phi+(E_M+E^\prime)c\phi]$ \\

$\Xi_c^0 \to \Xi^-\pi^+$
&$V_{cs}V_{ud}(T+E_{\bf B})$  \\
\hline
$\Xi_c^0 \to \Sigma^+ \pi^-$
&$\lambda_dE_M +\lambda_sE^\prime$\\
$\Xi_c^0 \to \Sigma^- \pi^+$
&$\lambda_d(T +E_{\bf B})$ \\
$\Xi_c^0 \to\Sigma( \Lambda)^0 \pi^0$
&$\frac{1}{\sqrt 2}[\lambda_d(-C-C^\prime-E_M+E_{\bf B})+\lambda_s(E_{\bf B}-E^\prime)]$      \\
$\Xi_c^0 \to \Sigma( \Lambda)^0\eta$
&$\frac{1}{\sqrt 2}[\lambda_d(C+C^\prime+E_M+E_{\bf B})+\lambda_s(E_{\bf B}+E^\prime)]c\phi$\\
&$-[\lambda_dE^\prime +\lambda_s(C+C^\prime+E_M)]s\phi$\\
$\Xi_c^0 \to \Sigma( \Lambda)^0\eta^\prime $
&$\frac{1}{\sqrt 2}[\lambda_d(C+C^\prime+ E_{\bf B}+E_M)+\lambda_s(E_{\bf B}+E^\prime)]s\phi$\\
&$+[\lambda_dE^\prime+\lambda_s(C+C^\prime+E_M)]c\phi$ \\
$\Xi_c^0 \to \Xi^-K^+$
&$\lambda_dE_{\bf B} +\lambda_s(T+E_{\bf B})$    \\
$\Xi_c^0 \to \Xi^0 K^0$
&$\lambda_dE_M +\lambda_s(C^\prime+E^\prime)$\\
$\Xi_c^0 \to p K^- $
&$\lambda_dE^\prime +\lambda_sE_M $ \\
$\Xi_c^0 \to n \bar{K}^0$
&$\lambda_d(C^\prime+E^\prime)+\lambda_sE_M$      \\
\hline
$\Xi_c^0 \to p \pi^-$
&$V_{cd}V_{us}(E_M+ E^\prime)$  \\
$\Xi_c^0 \to \Sigma^- K^+$
&$V_{cd}V_{us}(T+E_{\bf B})$  \\
$\Xi_c^0 \to \Sigma( \Lambda)^0 K^0$
&$V_{cd}V_{us}(C+C^\prime+E_M+E^\prime)$  \\
$\Xi_c^0 \to n \pi^0$
&$V_{cd}V_{us}\frac{1}{\sqrt 2}(E_{\bf B}-E_M -E^\prime)$  \\
$\Xi_c^0 \to n \eta $
&$V_{cd}V_{us}\frac{1}{\sqrt 2}(E_{\bf B}+E_M +E^\prime)c\phi$
$-V_{cd}V_{us}C^\prime s\phi$ \\
$\Xi_c^0 \to n \eta^\prime $
&$V_{cd}V_{us}\frac{1}{\sqrt 2}(E_{\bf B}+E_M+ E^\prime)s\phi$
$+V_{cd}V_{us}C^\prime c\phi$ \\
\rule{0pt}{16pt}
~&~\\
~&~\\
~&~\\
~&~\\
~&~\\
~&~\\
\hline
\end{tabular}
\begin{tabular}{|l|l|}
\hline
Decay modes& Amplitudes/$(\frac{G_F}{\sqrt 2})$ \\
\hline
$\Xi_c^+ \to \Sigma^+ \bar{K}^0$
&$V_{cs}V_{ud}(C+C^\prime)$ \\
$\Xi_c^+ \to \Xi^0 \pi^+$
&$V_{cs}V_{ud}(T+C^\prime)$\\
\hline
$\Xi_c^+ \to \Sigma( \Lambda)^0\pi^+$
&$\lambda_s(E_{\bf B}+E^\prime)+\lambda_d(T+C^\prime)$\\
$\Xi_c^+ \to \Sigma^+ \pi^0$
&$\frac{1}{\sqrt 2}[\lambda_s(E_{\bf B}+E^\prime)-\lambda_d C]$\\
$\Xi_c^+ \to \Sigma^+ \eta$
&$\frac{1}{\sqrt 2}[\lambda_dC+\lambda_s(E_{\bf B}+E^\prime)]c\phi-\lambda_s(C+C^\prime+E_M)s\phi$\\
$\Xi_c^+ \to \Sigma^+\eta^\prime $
&$\frac{1}{\sqrt 2}[\lambda_dC+\lambda_s(E_{\bf B}+E^\prime)]s\phi+\lambda_s(C+C^\prime+E_M)c\phi $ \\
$\Xi_c^+ \to \Xi^0 K^+$
&$\lambda_s(T+C^\prime+E_{\bf B}+E^\prime)$ \\
$\Xi_c^+ \to p \bar{K}^0$
&$\lambda_dC^\prime+\lambda_sE_M$
\\
\hline
$\Xi_c^+ \to \Sigma( \Lambda)^0 K^+$
&$V_{cd}V_{us}(T+C^\prime+E_{\bf B}+E^\prime)$ \\
$\Xi_c^+ \to \Sigma^+ K^0$
&$V_{cd}V_{us}(C+E_M)$\\
$\Xi_c^+ \to p \pi^0$
&$V_{cd}V_{us}\frac{1}{\sqrt 2}(E_{\bf B} +E^\prime-E_M)$\\
$\Xi_c^+ \to p \eta $
&$V_{cd}V_{us}\frac{1}{\sqrt 2}(E_{\bf B}+E_M+ E^\prime)c\phi
-V_{cd}V_{us}C^\prime s\phi $\\
$\Xi_c^+ \to p \eta^\prime$
&$V_{cd}V_{us}\frac{1}{\sqrt 2}(E_{\bf B}+E_M+ E^\prime)s\phi
+V_{cd}V_{us}C^\prime c\phi$ \\
$\Xi_c^+ \to n\pi^+$
&$V_{cd}V_{us}(E_{\bf B}+E^\prime)$ \\
\hline
\hline
Decay modes& Amplitudes/$(\frac{G_F}{\sqrt 2})$ \\
\hline
$\Lambda_c^+ \to \Sigma( \Lambda)^0\pi^+$
&$V_{cs}V_{ud}(T+C^\prime +E_{\bf B}+E^\prime)$\\
$\Lambda_c^+ \to \Sigma^+ \pi^0$
&$V_{cs}V_{ud}\frac{1}{\sqrt 2}(-C^\prime +E_{\bf B}+E^\prime)$ \\
$\Lambda_c^+ \to \Sigma^+ \eta$
&$V_{cs}V_{ud}[\frac{1}{\sqrt 2}(C^\prime+E_{\bf B}+E^\prime)c\phi- E_Ms\phi$]      \\
$\Lambda_c^+ \to \Sigma^+ \eta^\prime$
&$V_{cs}V_{ud}[\frac{1}{\sqrt 2}(C^\prime+E_{\bf B}+E^\prime)s\phi+ E_Mc\phi$]      \\
$\Lambda_c^+ \to \Xi^0 K^+$
&$V_{cs}V_{ud}(E_{\bf B}+E^\prime)$\\

$\Lambda_c^+ \to p \bar K^0$
&$V_{cs}V_{ud}(C+E_M)$ \\
\hline
$\Lambda_c^+ \to \Sigma^+ K^0$
&$\lambda_dE_M+\lambda_sC^\prime$\\
$\Lambda_c^+ \to \Sigma(\Lambda)^0 K^+$
&$\lambda_d(E_{\bf B}+E^\prime)+\lambda_s(T+C^\prime)$\\
$\Lambda_c^+ \to p \pi^0$
&$\lambda_d\frac{1}{\sqrt 2}(-C-C^\prime-E_M+E_{\bf B}+E^\prime)$ \\
$\Lambda_c^+ \to p \eta$
&$\lambda_d\frac{1}{\sqrt 2}(C+C^\prime+E_M+E_{\bf B}+E^\prime)c\phi-\lambda_s Cs\phi$      \\
$\Lambda_c^+ \to p \eta^\prime$
&$\lambda_d\frac{1}{\sqrt 2}(C+C^\prime+E_M+E_{\bf B}+E^\prime)s\phi+\lambda_s Cc\phi$    \\
$\Lambda_c^+ \to n \pi^+$
&$\lambda_d(T+C^\prime+E_{\bf B}+E^\prime)$\\
\hline
$\Lambda_c^+ \to p K^0$
&$V_{cd}V_{us}(C+C^\prime)$  \\
$\Lambda_c^+ \to n  K^+$
&$V_{cd}V_{us}(T+C^\prime)$    \\
\hline
\end{tabular}
\end{table}

\section{Numerical Results}
In the numerical analysis,
we perform a minimum $\chi^2$-fit,
with the equation written as~\cite{Geng:2018bow}
\begin{eqnarray}
\chi^2=
\sum_{i} \bigg(\frac{{\cal B}^i_{th}-{\cal B}^i_{ex}}{\sigma_{ex}^i}\bigg)^2+
\sum_{j}\bigg(\frac{{\cal R}^j_{th}-{\cal R}^j_{ex}}{\sigma_{ex}^j}\bigg)^2\,,
\end{eqnarray}
where ${\cal B}$ (${\cal R}$) denotes (the ratios of) the branching ratios.
The subscripts $th$ and $ex$ stand for
the theoretical inputs from the amplitudes in Table~\ref{tab1}
and the experimental data in Table~\ref{tab2},
respectively, with $\sigma_{ex}^{i,j}$ the experimental uncertainties.
By putting the recent observation of
${\cal B}(\Xi_c^0 \to \Xi^-\pi^+)=(1.80 \pm 0.55)\times 10^{-2}$~\cite{Li:2018qak}
into ${\cal R}_{1,2}(\Xi_c^0)$ in Table~\ref{tab2}, we determine that
${\cal B}(\Xi_c^0 \to \Xi^- K^+)=(5.0 \pm 1.9)\times 10^{-4}$ and
${\cal B}(\Xi_c^0 \to \Lambda^0\bar K^0)=(7.6 \pm 2.6)\times 10^{-3}$.
The ${\cal B}(\Xi_c^+ \to \Xi^0\pi^+)$ is extracted from the ratio of
${\cal B}(\Xi_c^+ \to \Xi^0\pi^+)/{\cal B}(\Xi_c^+\to\Xi^-\pi^+\pi^+)
=0.55\pm 0.16$~\cite{pdg} and the newly observed
${\cal B}(\Xi_c^+\to\Xi^-\pi^+\pi^+)
=(2.86\pm 1.21\pm 0.38)\times 10^{-2}$~\cite{Li:2019atu}.
We adopt the CKM matrix elements
in the Wolfenstein parameterization,
given by~\cite{pdg}
\begin{eqnarray}\label{B1}
&&(V_{cs},V_{ud},V_{us},V_{cd})=(1-\lambda^2/2,1-\lambda^2/2,\lambda,-\lambda)\,,
\end{eqnarray}
with $\lambda=s_c=0.22453\pm 0.00044$.
%
%
\begin{table}[t!]
\caption{The data for ${\bf B}_c\to {\bf B} M$.}\label{tab2}
{\footnotesize
\begin{tabular}{|c|c|}
\hline
Branching ratios
&Data
\\
\hline
$10^2{\cal B}(\Lambda_c^+ \to p \bar K^0)$
&$3.16\pm 0.16$~\cite{pdg}
\\
$10^2{\cal B}(\Lambda_c^+ \to \Lambda^0 \pi^+)$
&$1.30\pm0.07$~\cite{pdg}
\\
$10^2{\cal B}(\Lambda_c^+ \to \Sigma^0 \pi^+)$
&$1.29\pm 0.07$~\cite{pdg}
\\
$10^2{\cal B}(\Lambda_c^+ \to \Sigma^+ \pi^0)$
&$1.24\pm 0.10$~\cite{pdg}
\\
$10^2{\cal B}(\Lambda_c^+ \to \Xi^0 K^+)$
&$0.59\pm 0.09$~\cite{Ablikim:2018bir}
\\
$10^2{\cal B}(\Lambda_c^+ \to \Sigma^+ \eta)$
&$0.41\pm 0.20$~\cite{Ablikim:2018czr}
\\
$10^2{\cal B}(\Lambda_c^+ \to \Sigma^+ \eta^\prime)$
&$1.34\pm 0.57$~\cite{Ablikim:2018czr}\\[8mm]
\hline
\end{tabular}
\begin{tabular}{|c|c|}
\hline
(Ratios of) Branching ratios&Data
\\
\hline

$10^4{\cal B}(\Lambda_c^+ \to p \pi^0)$
&$0.8\pm1.4$ $(<0.27)$~\cite{Ablikim:2017ors,ppi0}
\\
$10^4{\cal B}(\Lambda_c^+ \to \Lambda^0 K^+)$
&$6.1\pm 1.2$~\cite{pdg}
\\
$10^4{\cal B}(\Lambda_c^+ \to \Sigma^0 K^+)$
&$5.2\pm 0.8$~\cite{pdg}
\\
$10^4{\cal B}(\Lambda_c^+ \to p \eta)$
&$12.4\pm 3.0$~\cite{pdg}
\\
$10^2{\cal B}(\Xi_c^0 \to \Xi^-\pi^+)$
&$1.80 \pm 0.55$~\cite{Li:2018qak}
\\
${\cal R}_1(\Xi_c^0)\equiv
\frac{{\cal B}(\Xi_c^0 \to \Xi^-K^+)}{{\cal B}(\Xi_c^0 \to \Xi^-\pi^+)}$
&$(0.56\pm0.12)s^2_c$~\cite{pdg}
\\
${\cal R}_2(\Xi_c^0)\equiv
\frac{{\cal B}(\Xi_c^0 \to \Lambda^0\bar K^0)}{{\cal B}(\Xi_c^0 \to \Xi^-\pi^+)}$
&$0.42\pm0.06$~\cite{pdg}
\\
$10^2{\cal B}(\Xi_c^+ \to \Xi^0\pi^+)$
&$1.57 \pm 0.84$~\cite{pdg,Li:2019atu}
\\
\hline
\end{tabular}}
\end{table}
%
Subsequently, we fit that
\begin{eqnarray}\label{su3_fit}
(T,C,C')&=&
(0.41\pm 0.02,0.47\pm 0.08,0.25\pm 0.02)\,\text{GeV}^3\,,\nonumber\\
(E_{\bf B},E_{M},E')&=&
(0.43\pm 0.04,0.14\pm 0.03,0.14\pm 0.07)\,\text{GeV}^3\,,\nonumber\\
(\delta_C,\delta_{C'},
\delta_{E_{\bf B}},\delta_{E_M},\delta_{E'})&=&
(31.3\pm 9.6, 158.1\pm 5.3, -81.0\pm 8.8, -32.8\pm 21.6, -88.0\pm 1.9)^\circ\,,\nonumber\\
\chi^2/n.d.f&=&0.5,
\end{eqnarray}
with $n.d.f=4$ as the number of degrees of freedom,
by which we present the branching ratios
of the ${\bf B}_c\to{\bf B}M$ decays in Table~\ref{tab_result},
together with the recent theoretical results for comparison.
%
\begin{table}[t!]
\caption{The numerical results of
the ${\bf B}_{c}\to {\bf B} M$ decays
with ${\cal B}_{{\bf B}M}\equiv {\cal B}({\bf B}_c\to {\bf B}M)$,
in comparison with the results from the $SU(3)_f$ symmetry~\cite{Geng:2019xbo,Zou:2019kzq}
and the calculation with the pole model,
current algebra and MIT bag model~\cite{Cheng:2018hwl,Zou:2019kzq}.
The data of ${\cal B}(\Xi_c^0\to \Xi^-K^+,\Lambda^0 \bar K^0)$
are extracted with $R_{1,2}(\Xi_c^0)$, respectively.}\label{tab_result}
{
\scriptsize
\begin{tabular}{|c|ccc|c|}
\hline
$\Xi_c^0$&$SU(3)_f$&Cheng~{\it et al.}&Our work&Expt.
\\
\hline
$10^3{\cal B}_{\Sigma^{+} K^{-}}$
&$7.6 \pm 1.4$
&7.8
&$22.0\pm5.7$
&\\
$10^3{\cal B}_{\Sigma^{0} \bar{K}^{0}}$
&$0.9^{+1.1}_{-0.9}$
&0.4
&$7.9 \pm 4.8$
&\\
$10^3{\cal B}_{\Xi^{0} \pi^{0}}$
&$10.0 \pm 1.4$
&18.2
&$4.7 \pm 0.9$
&\\
$10^3{\cal B}_{\Xi^{0} \eta}$
&$13.0\pm 2.3$
&26.7
&$8.3 \pm 2.3$
&\\
$10^3{\cal B}_{\Xi^{0} \eta'}$
&$9.1 \pm 4.1$
&
&$7.2 \pm 1.9$
&\\
$10^3{\cal B}_{\Xi^{-} \pi^{+}}$
&$29.5 \pm 1.4$
&64.7
&$19.3 \pm 2.8$
&$18.0 \pm 5.5$\\
$10^3{\cal B}_{\Lambda^{0} \bar{K}^{0}}$
&$14.2 \pm 0.09$
&13.3
&$8.3 \pm 5.0$
&$7.6\pm2.6$\\
\hline
$10^4{\cal B}_{\Sigma^{+} \pi^{-}}$
&$4.9 \pm 0.9$
&7.1
&$2.4 \pm 1.5$
&\\
$10^4{\cal B}_{\Sigma^{-} \pi^{+}}$
&$18.3 \pm 0.9$
&26.2
&$11.1\pm 1.6$
&\\
$10^4{\cal B}_{\Sigma^{0} \pi^{0}}$
&$5.0 \pm 0.9$&3.8
&$1.0\pm 0.5$
&\\
$10^4{\cal B}_{\Sigma^{0} \eta}$
&$1.8\pm1.1$&0.5
&$2.5 \pm 1.3$
&\\
$10^4{\cal B}_{\Sigma^{0} \eta'}$
&$3.2\pm2.2$&
&$0.1 \pm 0.1$
&\\
$10^4{\cal B}_{\Xi^{-} K^{+}}$
&$12.8 \pm 0.6$&39.0
&$5.6 \pm 0.8$
&$5.0\pm1.9$\\
$10^4{\cal B}_{\Xi^{0} K^{0}}$
&$9.6 \pm 0.4$&13.2
&$6.3\pm 1.9$
&\\
$10^4{\cal B}_{p K^{-}}$
&$6.0 \pm 1.3$&3.5
&$2.5 \pm 1.6$
&\\
$10^4{\cal B}_{n \bar{K}^{0}}$
&$10.7 \pm 0.6$&14.0
&$7.8 \pm 2.3$
&\\
$10^4{\cal B}_{\Lambda^{0} \pi^{0}}$
&$3.1\pm 1.1$&2.4
&$1.0 \pm 0.5$
&\\
$10^4{\cal B}_{\Lambda^{0} \eta}$
&$7.9\pm2.7$&7.7
&$2.6 \pm 1.3$
&\\
$10^4{\cal B}_{\Lambda^{0} \eta'}$
&$16.4\pm10.6$&
&$0.2 \pm 0.1$&\\
\hline
$10^5{\cal B}_{p \pi^{-}}$
&$3.1 \pm 0.7$&
&$7.6 \pm 2.0$&\\
$10^5{\cal B}_{\Sigma^{-} K^{+}}$
&$6.1 \pm 0.4$&
&$5.5 \pm 0.7$&\\
$10^5{\cal B}_{\Sigma^{0} K^{0}}$
&$2.5 \pm 0.2$&
&$2.3 \pm 1.4$&\\
$10^5{\cal B}_{n \pi^{0}}$
&$1.5 \pm 0.4$&
&$3.3 \pm 0.9$&\\
$10^5{\cal B}_{n \eta}$
&$5.2\pm 2.1$&
&$4.2\pm 0.8$&\\
$10^5{\cal B}_{n \eta'}$
&$ 10.2\pm7.1$&
&$1.2\pm0.3$&\\
$10^5{\cal B}_{\Lambda^{0} K^{0}}$
&$0.6 \pm 0.2$&
&$2.4\pm1.4$&\\[36.5mm]
\hline
\end{tabular}
\begin{tabular}{|c|ccc|c|}
\hline
$\Xi_c^+$&$SU(3)_f$&Cheng~{\it et al.}&Our work&Expt.
\\
\hline
$10^3{\cal B}_{\Sigma^{+} \bar{K}^{0}}$
&$7.8^{+10.2}_{-\;\,7.8}$&2.0
&$24.1\pm 7.1$&\\
$10^3{\cal B}_{\Xi^{0} \pi^{+}}$
&$4.2 \pm 1.7 $&17.2
&$9.3 \pm 3.6$&$15.7\pm 8.4$\\ 
\hline
$10^4{\cal B}_{\Sigma^{0} \pi^{+}}$
&$26.5 \pm 2.5$&43.0
&$13.4 \pm 4.9$&\\
$10^4{\cal B}_{\Sigma^{+} \pi^{0}}$
&$26.1 \pm 6.7$&13.6
&$16.4 \pm 3.2$&\\
$10^4{\cal B}_{\Sigma^{+} \eta}$
&$15.0\pm 10.6$&3.2
&$14.1\pm3.9$&\\
$10^4{\cal B}_{\Sigma^{+} \eta'}$
&$34.6\pm21.9$&
&$8.7\pm 3.7$&\\
$10^4{\cal B}_{\Xi^{0} K^{+}}$
&$7.6 \pm 1.6$&22.0
&$12.3 \pm3.1$&\\
$10^4{\cal B}_{p \bar{K}^{0}}$
&$46.4 \pm 7.2$&39.6
&$48.6\pm 12.2$&\\
$10^4{\cal B}_{\Lambda^{0} \pi^{+}}$
&$12.3 \pm 4.2$&8.5
&$13.9 \pm 5.1$&\\ 
\hline
$10^5{\cal B}_{\Sigma^{0} K^{+}}$
&$11.9 \pm 0.7$&
&$7.2\pm 1.8$&\\
$10^5{\cal B}_{\Sigma^{+} K^{0}}$
&$19.5 \pm 1.7$&
&$16.9\pm 5.4$&\\
$10^5{\cal B}_{p \pi^{0}}$
&$6.0 \pm 1.4$&
&$1.5\pm1.5$&\\
$10^5{\cal B}_{p \eta}$
&$20.4\pm8.4$&
&$16.6\pm 3.1$&\\
$10^5{\cal B}_{p \eta'}$
&$40.1\pm27.7$&
&$5.8\pm 1.5$&\\
$10^5{\cal B}_{n \pi^{+}}$
&$12.1 \pm 2.8$&
&$5.2\pm 1.5$&\\
$10^5{\cal B}_{\Lambda^{0} K^{+}}$
&$3.3 \pm 0.8$&
&$7.5\pm 1.9$&\\
\hline
%
%
\hline
$\Lambda_c^+$&$SU(3)_f$&Cheng~{\it et al.}&Our work&Expt.
\\
\hline
$10^2{\cal B}_{\Sigma^{0} \pi^{+}}$
&$1.26 \pm 0.06$&2.24
&$1.26 \pm0.32$&$1.29\pm0.07$\\
$10^2{\cal B}_{\Sigma^{+} \pi^{0}}$
&$1.26 \pm 0.06$&2.24
&$1.23 \pm 0.17$  &$1.24\pm0.10$\\
$10^2{\cal B}_{\Sigma^{+} \eta}$
&$0.29\pm0.12$&0.74
&$0.47 \pm 0.22$&$0.41\pm0.20$\\
$10^2{\cal B}_{\Sigma^{+} \eta'}$
&$1.44\pm0.56$&
&$0.93 \pm 0.28$&$1.34\pm0.57$\\
$10^2{\cal B}_{\Xi^{0} K^{+}}$
&$0.57\pm0.09$&0.73
&$0.59 \pm 0.17$ &$0.59\pm0.09$\\
$10^2{\cal B}_{p \bar{K}^{0}}$
&$3.14 \pm 0.15$&2.11
&$3.14 \pm 1.00$&$3.16\pm0.16$\\
$10^2{\cal B}_{\Lambda^{0} \pi^{+}}$
&$1.27\pm0.07$&1.30
&$1.32 \pm 0.34$ &$1.30\pm0.07$\\
\hline
$10^4{\cal B}_{\Sigma^{+} K^{0}}$
&$10.5\pm 1.4$&14.4
&$19.1 \pm 4.8$&\\
$10^4{\cal B}_{\Sigma^{0} K^{+}}$
&$5.2 \pm 0.7$&7.2
&$5.5 \pm 1.6$ &$5.2\pm0.8$\\
$10^4{\cal B}_{p \pi^{0}}$
&$1.1^{+1.3}_{-1.1}$&1.3
&$0.8 ^{+0.9}_{-0.8}$ &$0.8\pm1.4$\\
$10^4{\cal B}_{p \eta}$
&$11.2\pm2.8$&12.8
&$11.4\pm 3.5$ &$12.4\pm3.0$\\
$10^4{\cal B}_{p \eta'}$
&$24.5\pm14.6$&
&$7.1 \pm 1.4$ &\\
$10^4{\cal B}_{n \pi^{+}}$
&$7.6\pm 1.1$&0.9
&$7.7\pm 2.0$&\\
$10^4{\cal B}_{\Lambda^{0} K^{+}}$
&$6.6\pm0.9$&10.7
&$5.9 \pm 1.7$&$6.1\pm1.2$\\ 
\hline
$10^5{\cal B}_{p K^{0}}$
&$1.2^{+1.4}_{-1.2}$&
&$3.7\pm 1.1$&\\
$10^5{\cal B}_{n K^{+}}$
&$0.4\pm0.2$&
&$1.4\pm 0.5$&\\ 
\hline
\end{tabular}
}
\end{table}

\section{Discussions and Conclusions}

Since $\chi^2/n.d.f=0.5$ presents a good fit,
we demonstrate that
the diagrammatic approach can explain the data well.
Moreover,
the non-factorizable effects depicted in Fig.~\ref{fig1} now have clear information.
The fit of $|E_B|\simeq (|T|,|C|)\simeq 0.4$ and
$|C'|\simeq 2(|E_M|,|E'|)\simeq 0.3$
shows that the non-factorizable effects
can be as significant as the factorizable ones;
nonetheless, neglected in the factorization approach.
While the theoretical computations are unable to
explain ${\cal B}(\Lambda_c^+ \to \Xi^0 K^+)$~\cite{XiK_1,XiK_2,XiK_3,XiK_4,XiK_5},
we explicitly present
the two $W$-exchange effects as $E_{\bf B}$ and $E^\prime$
that contribute to $\Lambda_c^+ \to \Xi^0 K^+$,
together with the relative phases.
Besides, we point out that $E_{\bf B}$ has the main contribution.
The $E_M$ term as the rarely studied $W$-exchange process
is found to have a significant interference with $C$ in $\Lambda_c^+ \to p \bar K^0$.
In contrast, the $SU(3)_f$ parameters cannot distinguish
the three $W$-exchange contributions.
As seen in Table~\ref{tab1},
the $\Lambda_c^+ \to \Sigma^+ \eta^{(\prime)},\Sigma^+ K^0$ decays
only receive the non-factorizable effects.
Particularly,
$C'$ and $E_M$ in $\Lambda_c^+ \to\Sigma^+ K^0$
give a constructive interference,
leading to ${\cal B}(\Lambda_c^+ \to \Sigma^+ K^0)=(19.1\pm 4.8)\times 10^{-4}$
accessible to the BESIII experiment.
We hence have a better understanding for the non-factorizable effects.

In the $SU(3)_f$ symmetry,
the ${\cal B}(\Lambda_c^+\to p\pi^0)$ was once overestimated as
two times larger than the experimental upper bound~\cite{Geng:2017esc}.
By recovering one of the previously neglected parameters,
which gives the destructive interference,
the number has been reduced to agree with the data~\cite{Geng:2018rse,Geng:2019xbo}.
It is interesting to note that the recovered parameter is recognized as
a factorizable effect, which corresponds to our $C$ term in $\Lambda_c^+\to p\pi^0$.
The $SU(3)_f$ symmetry derives that
${\cal A}(\Xi_c^0\to \Xi^- \pi^+)=
s_c {\cal A}(\Xi_c^0\to \Xi^- K^+)$~\cite{Geng:2017esc,Geng:2018plk,He:2018joe}.
This causes ${\cal R}_1(\Xi_c^0)\equiv
{\cal B}(\Xi_c^0 \to \Xi^-K^+)/{\cal B}(\Xi_c^0 \to \Xi^-\pi^+)\simeq 1.0s_c^2$
to be 4$\sigma$ away from the observation.
%
\begin{figure}
\centering
\includegraphics[width=0.9\textwidth]{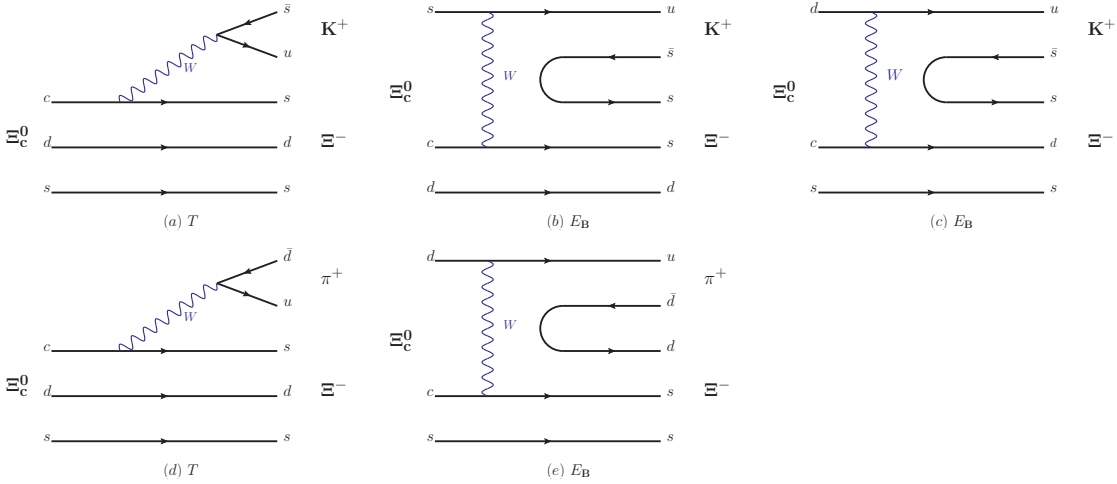} \\
\caption{Decay processes (a,b,c) and (d,e)
for $\Xi_c^0\to \Xi^- K^+$ and $\Xi_c^0\to \Xi^- \pi^+$,
respectively.}\label{fig2}
\end{figure}
%
According to the topological diagrams in Fig.~\ref{fig2},
we find that
\begin{eqnarray}\label{eq_R1}
&&\bar {\cal A}_K\equiv {\cal A}(\Xi_c^+\to \Xi^- K^+)/(\frac{G_F}{\sqrt 2})
=V_{cs}V_{us}T+V_{cs}V_{us}E_{\bf B}+V_{cd}V_{ud}E_{\bf B}\,,\nonumber\\
&&\bar {\cal A}_{\pi}\equiv {\cal A}(\Xi_c^+\to \Xi^- \pi^+)/(\frac{G_F}{\sqrt 2})
=V_{cs}V_{ud}(T+E_{\bf B})\,,
\end{eqnarray}
where $\bar {\cal A}_K$ is reduced as $\bar {\cal A}_K=V_{cs}V_{us}T$
with $V_{cs}V_{us}=-V_{cd}V_{ud}$,
resulting in
${\cal R}_1(\Xi_c^0)\simeq (1+E_{\bf B}/T)^2 s_c^2$
with the value of $(0.54\pm0.04)s^2_c$ to agree with the data.
Indeed,
the effects of the $SU(3)_f$ symmetry breaking
can give rise to the new parameters added to $\Xi_c^0 \to \Xi^-K^+$,
instead of $\Xi_c^0\to \Xi^- \pi^+$,
such that ${\cal R}_1(\Xi_c^0)$ can be fit~\cite{Geng:2018bow}.
Here, our interpretation for ${\cal R}_1(\Xi_c^0)$ relies on 
the additional diagram in Fig.~\ref{fig2}
without invoking the $SU(3)_f$ symmetry breaking.

By relating the topological diagrams in Fig.~\ref{fig1} to
the symmetry properties of the baryon wave functions,
such as the (anti-)symmetric quark ordering of $\Sigma^0(\Lambda^0)\sim (ud\pm du)s$
or the irreducible forms in the $SU(3)_f$ and $SU(2)$ spin symmetries,
one can derive the more restrict parameterization of
the topological amplitudes~\cite{Kohara:1991ug,Chau:1995gk}.
This leads to ${\cal R}_1(\Xi_c^0)\simeq 1.0s_c^2$ inconsistent with the data.
Moreover, the $T$ term, which contributes to
$\Lambda_c^+ \to \Sigma^0 M^+$ with $M^+=(\pi^+,K^+)$ in Table~\ref{tab1},
becomes forbidden in Refs.~\cite{Kohara:1991ug,Chau:1995gk}.
We hence turn off the $T$ term in $\Lambda_c^+ \to \Sigma^0 M^+$
as a test fit, and obtain $\chi^2\sim 30$.
Clearly,
the more restrict representations cannot explain the data well.
Without considering the symmetry properties of the baryon wave functions,
our topological amplitudes present the most general forms,
which are able to receive the short and long-distance contributions both.
The long-distance effect has been proposed in the pole model
to contribute to the $W$-exchange process~\cite{Cheng:2018hwl,Zou:2019kzq}.
In the $\Xi_c^0\to \Xi^- M^+$ decay,
$\Xi_c^0$ transforms as the $\Lambda^0$ ($\Sigma^0$) pole
followed by the strong decay $\Lambda^0(\Sigma^0)\to\Xi^- M^+$,
which contributes to Figs.~\ref{fig2}b and c; nonetheless,
the latter diagram is forbidden
with the more restrict representations.
Another example comes from the rescattering effect.
With $\Lambda^*$ denoting the higher-wave $\Lambda$ state,
the $\Lambda_c^+\to \Lambda^* \rho^+$ decay has a $T$ amplitude ($T^*$).
Through the $\pi^0$ exchange, $\Lambda^* \rho^+$
rescatter into $\Sigma^0 \pi^+$, such that
$T^*$ contributes to $T$ in $\Lambda_c^+\to \Sigma^0 \pi^+$.
In fact, the symmetry properties of
the meson wave functions are not involved in the $D\to MM$ decays,
such that the long-distance contributions have been
absorbed into the topological amplitudes~\cite{Cheng:2012xb,Li:2012cfa,
Cheng:2019ggx,Hsiao:2019ait}.

In Table~\ref{tab_result},
the $SU(3)_f$ parameters and topological amplitudes
that respect the $SU(3)_f$ symmetry are found to explain
the data of ${\cal B}(\Lambda_c^+\to {\bf B}M)$ well,
indicating that
the $SU(3)_f$ symmetry in $\Lambda_c^+\to {\bf B}M$
has no sizeable broken effects.
In the $D\to K\bar K$ decays,
the $W$-exchange processes with $V_{cs}V_{us}$ and $V_{cd}V_{ud}$ are
parameterized as $E^{(d)}$ and $E^{(s)}$, respectively.
One needs the sizeable broken effect of $|E^{(s)}|>|E^{(d)}|$
to explain ${\cal B}(D^0\to K^+ K^-)/{\cal B}(D^0\to \pi^+ \pi^-)$ and
${\cal B}(D^0\to K^0\bar K^0)$~\cite{Cheng:2012xb,Li:2012cfa,Li:2013xsa,Cheng:2019ggx}.
Likewise, since we can present
${\cal A}(\Xi_c^0\to \Xi^- K^+)$ as
$\bar {\cal A}_K\propto T+\Delta E_{\bf B}$
with $\Delta E_{\bf B}\equiv E_{\bf B}^{(s)}-E_{\bf B}^{(d)}$, Eq.~(\ref{eq_R1}),
whether $\Delta E_{\bf B}$ is equal to zero or not
can be a test of the broken $SU(3)_f$ symmetry.
This requires more accurate measurements
from BELLEII and LHCb.

We compare
the three theoretical results in Table~\ref{tab_result},
which all agree with the observed ${\cal B}(\Lambda_c^+\to {\bf B}M)$.
However,
the $SU(3)_f$ symmetry gives ${\cal B}(\Xi_c^+\to \Xi^0\pi^+)$
at least 2 times smaller than the observation~\cite{Geng:2019xbo,Zou:2019kzq}.
Moreover,
the $SU(3)_f$ symmetry and the computations
that involve the factorization, pole model, current algebra and
MIT bag model
present ${\cal B}(\Xi_c^0\to \Xi^-\pi^+,\Xi^- K^+,\Lambda^0\bar K^0)$
with large deviations from the data~\cite{Geng:2019xbo,Cheng:2018hwl,Zou:2019kzq}.
Since we are able to explain ${\cal B}(\Xi_c^+\to \Xi^0\pi^+)$ and
${\cal B}(\Xi_c^0\to \Xi^-\pi^+,\Xi^- K^+,\Lambda^0\bar K^0)$,
the topological amplitudes are shown to have the advantage
to explain the two-body $\Xi_c^{0,+}$ decays.
To further investigate the $\Xi_c^{0,+}\to {\bf B}M$ decays,
we present our predictions in Table~\ref{tab_result}.
It is worth noting that
$E_M$ and $E^\prime$ give a constructive (destructive) interference to
$\Xi_c^0 \to \Sigma^+ K^-(\pi^-)$, such that we obtain
${\cal B}(\Xi_c^0 \to \Sigma^+ K^-)=(22.0\pm5.7)\times 10^{-3}$
and
${\cal B}(\Xi_c^0 \to \Sigma^+ \pi^-)=(2.4 \pm 1.5)\times 10^{-4}$,
bigger and smaller than the other predictions, respectively.
In addition to $\Lambda_c^+\to p\pi^0$,
there are three singly Cabibbo-suppressed
$\Lambda_c^+$ decay modes to be measured.
Their branching ratios are predicted as
${\cal B}(\Lambda_c^+ \to n \pi^{+},p \eta^\prime,\Sigma^{+} K^{0})
=(7.7\pm 2.0,7.1 \pm 1.4,19.1 \pm 4.8)\times 10^{-4}$,
where ${\cal B}(\Lambda_c^+ \to n \pi^{+})$ is consistent with
the value from the $SU(3)_f$ symmetry, whereas ${\cal B}(\Lambda_c^+\to\Sigma^{+} K^{0})$
agrees with the theoretical computation in~\cite{Cheng:2018hwl,Zou:2019kzq}.

In summary, we have studied the ${\bf B}_c\to{\bf B} M$ decays
within the framework of the diagrammatic approach
that respects the $SU(3)_f$ symmetry.
With the extraction of the topological amplitudes,
we have explicitly presented
the two $W$-exchange effects as $E_{\bf B}$ and $E^\prime$
that contribute to the non-factorizable $\Lambda_c^+ \to \Xi^0 K^+$ decay,
together with the relative phases, where $E_{\bf B}$ gives the main contribution.
We have obtained
${\cal B}(\Lambda_c^+\to p\pi^0)=(0.8 ^{+0.9}_{-0.8})\times 10^{-4}$,
which agrees with the experimental upper bound.
We have presented that
${\cal B}(\Xi_c^+\to\Xi^{0} \pi^{+})=(9.3 \pm 3.6)\times 10^{-3}$,
${\cal B}(\Xi_c^0\to \Xi^-\pi^+,\Lambda^0\bar K^0)
=(19.3 \pm 2.8,8.3 \pm 5.0)\times 10^{-2}$ and
${\cal B}(\Xi_c^0\to \Xi^- K^+)=(5.6 \pm 0.8)\times 10^{-4}$
to agree with the data,
whereas the other theoretical results have shown sizeable deviations.
We have predicted that
${\cal B}(\Lambda_c^+ \to n \pi^{+},p \eta^\prime,\Sigma^{+} K^{0})
=(7.7\pm 2.0,7.1 \pm 1.4,19.1 \pm 4.8)\times 10^{-4}$,
in order to be compared with future BESIII, BELLEII and LHCb measurements.

\section*{ACKNOWLEDGMENTS}
We would like to thank Professor H.Y. Cheng,
Professor Fanrong Xu, and Professor C.Q.~Geng for useful discussions.
We would like to thank Dr.~Eduardo Rodrigues for
reading the manuscript and giving valuable comments.
This work was supported by
National Science Foundation of China (11675030) and
(11905023).

\end{document}